# Superconducting $Sn_{1-x}In_xTe$ Nanoplates

*Satoshi Sasaki and Yoichi Ando\**

Institute of Scientific and Industrial Research, Osaka University, Ibaraki, Osaka 567-0047, Japan

*E-mail: y_ando@sanken.osaka-u.ac.jp.

**ABSTRACT:** Recently, the search for Majorana fermions has become one of the most prominent subjects in condensed matter physics. This search involves explorations of new materials and hence offers interesting opportunities for chemistry. Theoretically, Majorana fermions may reside in various types of topological superconductor materials, and superconducting $Sn_{1-x}In_xTe$, which is a doped topological crystalline insulator, is one of the promising candidates to harbor Majorana fermions. Here, we report the first successful growth of superconducting $Sn_{1-x}In_xTe$ nanoplates on Si substrates by a simple vapor transport method without employing any catalyst. We observed robust superconducting transitions in those nanoplates after device fabrication and found that the relation between the critical temperature and the carrier density is consistent with that of bulk single crystals, suggesting that the superconducting properties of the nanoplate devices are essentially the same as those of bulk crystals. With the help of nanofabrication, those nanoplates would prove useful for elucidating the potentially topological nature of superconductivity in $Sn_{1-x}In_xTe$ to harbor Majorana fermions and thereby contribute to the future quantum technologies.

**INTRODUCTION**

Majorana fermions are theoretically predicted particles to possess a distinct property that particle is its own antiparticle.[1] They are expected to emerge as "quasiparticles" at the edge or surface of a special type of superconductors called topological superconductors (TSCs).[2-4] The search for TSCs is a hot topic in physics and offers interesting opportunities for chemistry. While topological superconducting states can be realized in many different systems,[5-12] degenerately-doped topological insulators or topological crystalline insulators[13-16] with strong spin-orbit coupling are considered to be a promising platform.[17-21] In this regard, the topological crystalline insulator SnTe doped with In, $Sn_{1-x}In_xTe$, is one of the candidates of TSCs. Possible existence of the surface Majorana fermions (and hence the occurrence of topological superconductivity) in bulk $Sn_{1-x}In_xTe$ samples was detected in 2012 by point-contact spectroscopy,[19] and successive studies elucidated its superconducting phase diagram with respect to the superconducting transition temperature $T_c$ and the carrier density,[22] or $T_c$ vs In content.[23,24] However, detailed studies of possible Majorana fermions using reliable tunnel junctions have not been performed, because of the difficulty in growing high-quality thin films of superconducting $Sn_{1-x}In_xTe$. Also, the nature of surface electronic states of superconducting $Sn_{1-x}In_xTe$ on different crystal faces remains unexplored, even though such a study would be crucial for elucidating a TSC.

In this regard, superconducting nanoplates which have relatively large top and bottom surfaces provide ideal settings for selectively accessing a targeted surface, using lithography and nanofabrication techniques. Also, to confirm the topological superconductivity in $Sn_{1-x}In_xTe$, it is of particular importance to fabricate advanced devices suitable for confirming peculiar properties of Majorana fermions.[3,4] Unfortunately, unlike graphene or $Bi_2Se_3$, bulk single crystals of $Sn_{1-x}In_xTe$, which have the rock salt structure without any van der Waals gap, are difficult to be cleaved into thin platelets with a flat surface, which is a prerequisite for any nanofabrication process. Nanoplates of superconducting $Sn_{1-x}In_xTe$ can solve this problem, and the development of a simple and inexpensive growth technique for such samples would greatly foster future studies of topological superconductivity.

In addition, the vapor-transport growth of nanoplates has several advantages in comparison with the growth of "large" bulk single crystals: First, better homogeneity and crystallinity are naturally expected because of their reduced size. Second, the nanometer-size of the crystals requires a drastically shorter growth time in comparison to the growth of large-size crystals by the same vapor-transport technique (e.g. 10 minutes vs 2 weeks).

Since the discovery of the topological crystalline insulator nature of SnTe in 2012 by Tanaka *et al.*,[25] this material is attracting significant attention as a new type of topological material. Several groups have reported syntheses of nanostructures of SnTe, the parent material of $Sn_{1-x}In_xTe$, using vapor-liquid-solid (VLS) and vapor-solid (VS) growth techniques with and without Au catalyst.[26-30] In particular, SnTe nanoplates with both {100} and {111} surfaces were successfully grown using Au catalyst, but without the catalyst, only nanoblocks of SnTe were reported to grow.[30] Here, we report the growth of $Sn_{1-x}In_xTe$ nanoplates with {100} and {111} top and bottom faceted surfaces [Figures 1d–g] using the VS growth technique with neither transport agent nor catalyst. An advantage of this simple growth method is that one can obtain clean samples without additional impurities induced during the growth process, which is good for superconductivity. Relatively large $Sn_{1-x}In_xTe$ nanoplates with lateral dimensions typically in the range of 10 micrometer are easily obtained with the present method. Their thickness is roughly correlated with the growth time and hence is tunable. Our $Sn_{1-x}In_xTe$ nanoplates with the In content of 6 - 11% exhibit robust superconductivity, which is confirmed by transport measurements of six-terminal devices fabricated from the nanoplates.

**EXPERIMENTAL SECTION**

**$Sn_{1-x}In_xTe$ Nanoplate Synthesis with neither Transport Agent nor Catalyst.** Single-crystalline nanoplates of $Sn_{1-x}In_xTe$ were grown by a vapor-transport method without transport agent. First, to synthesize a homogeneous polycrystalline source for the crystal growth, a stoichiometric mixture of high-purity elements of Sn (99.99%), In (99.99%), and Te (99.999%) was melted in a sealed evacuated quartz glass tube and was kept at 950 °C for 72 h with intermittent shaking to promote homogeneity. The obtained material was transferred to another quartz tube of a larger diameter together with a Si substrate of about 10 mm × 20 mm size under inert atmosphere. Then the quartz tube was sealed and put into a horizontal three-zone tube furnace; in the quartz tube, the Si substrate was placed next to the source material on the lower-temperature side as illustrated in Figure 1a. The growth was performed by first raising the temperature of the source material to 500 °C at the rate of 8 °C/min and then to 600 °C at the rate of 1.7 °C/min; after that, the source temperature was kept at 600 °C [$T_1$ as indicated in Figure 1a] for a specified time period. The Si substrate was placed in the temperature gradient ($\nabla T$) of 1.4 °C/cm. It is worth noting that the actual temperature of the Si substrate [$T_2$ as indicated in Figure 1a] can be lower than that estimated from the $\nabla T$ applied by the tube furnace under the quasi-adiabatic condition inside the evacuated quartz tube, where the Si substrate can be heated mainly by the radiation.[31] The typical duration of the crystal growth was 10 min. After the growth, the quartz tube was furnace-cooled to room temperature to avoid thermal shocks on the nanoplate samples.

**Details of the Experimental Procedures.** The nanoplate growth took place on a Si substrate placed inside a sealed evacuated quartz glass tube with the inner/outer diameter of 10 mm/12 mm and ~ 80 mm length using a three-zone horizontal tube furnace that can apply a 1.4 K/cm temperature gradient centered at ~600 °C near the polycrystalline $Sn_{1-x}In_xTe$ source material. The Si substrate was carefully washed by hot acetone (~ 65 °C) with ultrasonic agitation for 3 min to get rid of possible organic contaminations. Before weighing, an oxidized layer of Sn shots was removed by hydrogen reduction and the treated Sn shots were kept in either a good vacuum or inert Ar atmosphere to prevent re-oxidization. High purity elemental shots of Sn, In, and Te were mixed in the stoichiometric ratio and put into a quartz glass tube inside a glove box with Ar atmosphere. Then, the tube was evacuated down to $10^{-2}$ Pa and sealed by a torch for the synthesis of a homogeneous polycrystalline $Sn_{1-x}In_xTe$ source material at 950 °C. For the growth of nanoplates, the source material and the pre-washed Si substrate were transferred into a larger quartz glass tube in a glove box, and the tube was subsequently evacuated and sealed. The growth temperature was ~600 °C. The thickness of the nanoplates can be roughly controlled from ~ 40 nm to ~ 200 nm by varying the growth time period from ~10 min to ~40 min.

**$Sn_{1-x}In_xTe$ nanodevice fabrications.** The thickness of $Sn_{1-x}In_xTe$ nanoplates was measured by a laser microscope (Keyence VKX200) to select nanoplates thinner than 100 nm for the convenience of nanodevice fabrication. A standard PMMA (PolyMethyl Methacrylate) resist was used for EB lithography. In all lithography processes, the sample temperature was kept below 110 °C to ensure that the superconducting properties of nanoplates are not affected. The metal electrodes were made by thermal deposition of a 5-nm Pd buffer layer and successive thermal deposition of 65-nm Au film. After the thermal deposition, a 30-nm Pd film was deposited by RF sputtering to reinforce the connection between the deposited films on the nanoplate and on the Si substrate.

**RESULTS AND DISCUSSION**

Only two types of $Sn_{1-x}In_xTe$ nanoplates with either {100} or {111} faceted surfaces [schematically shown in Figures 1b,c] have been found on the Si substrates after the growth. The shape/size of typical samples is shown in Figures 1d–g. We found that the number of nanoplates grown with {100} surface (the maximum of ~20 nanoplates/cm$^2$) is much larger than that of nanoplates with {111} surface. Moreover, we have never observed nanoplates with {110} surface. This suggests that the surface energy for our growth conditions is the lowest for the growth along the [100] direction, becoming slightly higher for the [111] and much higher for the [110] directions, which is consistent with the conclusion on the preferential growth orientation in SnTe crystalline nanostructures reported in Ref. (29). It is worth noting that the surface energies have indeed been predicted to be determinants for particle morphology;[32] in this regard, SnTe nanostructures are a useful playground for nanomorphology.[33] We note that nanowires of SnTe were also found to grow horizontally on silicon substrates alongside the nanoplates in our experiments, but it is beyond the scope of this paper and we will elaborate on it elsewhere.

The above discussion on the growth orientation assumes that the ternary (Sn,In)Te system behave like binary SnTe. This is likely to be reasonable as long as the In content is low, which is actually the case here. We note that the InTe is not isostructural to SnTe at ambient pressure, and it has been reported that $Sn_{1-x}In_xTe$ keeps the same rock-salt structure as SnTe up to $x \sim 0.5$.[34]

A good homogeneity of grown samples and the absence of any segregation of constituent elements in the nanoplates have been confirmed by elemental mapping with Electron Probe

Microanalyzer (EPMA). Typical EPMA elemental maps of the vapor-grown (100) nanoplate [shown in Figure 2a] indicate that all the constituent elements are homogeneously distributed within the nanoplate sample as shown in Figures 2b–d. Further evidence for the high quality of grown nanoplates comes from the EPMA qualitative analysis, which probes the characteristic X-ray spectra of a wide range of chemical elements in investigated samples; as shown in Figure S1 in the Supporting Information, no peaks other than In, Sn, and Te have been observed in the measured EPMA spectra of our $Sn_{1-x}In_xTe$ nanoplates.

The analysis of the EPMA spectra can also give the indium content $x$, which is determined by averaging the values obtained from the EPMA quantitative analysis on three different points; since the In distribution is homogeneous, the error (mean square of the three values) is only 0.001−0.002. For the nanoplate shown in Figure 2, which was grown from polycrystalline $Sn_{0.85}In_{0.15}Te$, $x$ was found to be 0.070. The nanoplates fabricated into devices, which are shown in the insets of Figures 3b,d, were also grown from the same source and their In content $x$ was found to be 0.110 and 0.061, respectively. The relatively large $x$ values in the vapor-grown nanoplates in comparison to those of bulk single crystals grown on the inner wall of the quartz glass tube by a similar method[22] is remarkable. This difference may be understood as a result of a large temperature drop between the source material ($T_1$) and the nanoplates on the Si substrate ($T_2$); a large temperature difference corresponds to a large difference in the saturated vapor pressure (SVP), which is a driving force for the condensation of the relevant atoms and/or molecules, and the temperature dependence of the SVP was reported to be steeper for In atoms than for SnTe molecules [i.e. $\delta(-\log P(atm))/\delta(1000/T(K))$ was ~12 for In and ~10 for SnTe].[35,36] This may explain the increase of the In content in nanoplates grown on the Si substrate, for which the temperature $T_2$ is lower than the temperature of surrounding walls. The enhanced In content in comparison to bulk crystals[22] is another advantage of our growth method, since it leads to higher attainable $T_c$.

Having large and thin nanoplates of $Sn_{1-x}In_xTe$, we are able to fabricate Hall-bar devices for transport measurements using the electron beam (EB) lithography technique. Insets of Figures 3b,d show two devices made from nanoplates with the (100) surface, device A (57-nm thick) and device B (87-nm thick), both have lateral dimensions of several micrometers. The resistivity $\rho_{xx}$ and the Hall resistivity $\rho_{yx}$ have been measured in these devices as a function of temperature $T$ and magnetic field $B$ using a six-probe method and a standard lock-in technique. The Quantum Design Physical Properties Measurement System was used as a platform to cool the samples down to 0.34 K and apply magnetic fields up to 9 T.

Our key observation in both samples is robust superconductivity. Figures 3a,c show $\rho_{xx}(T)$ of devices A and B, respectively. A sharp superconducting transition was observed at 2.20 K (mid-point) for the device A with $x = 0.110$ and at 1.75 K for the device B with $x = 0.061$, suggesting that the hole density in these samples is different. For both devices, the residual resistivity at 4 K, $\rho_{4K}$, was ~ 0.6 mΩcm which is similar to the values obtained in bulk crystals.[22] A linear magnetic-field dependence of $\rho_{yx}$ in both devices [shown in Figures 3b,d] indicates that only a single band with p-type carriers dominates the transport in both samples. From the slope of $\rho_{yx}(B)$ measured at 5 K [Figures 3b,d], the nominal carrier density $p = r/(eR_H)$, where $e$ is the electron charge, $R_H$ is the Hall coefficient, and $r$ is the Hall factor, is $1.5 \times 10^{21}$ cm$^{-3}$ (device A) and $0.9 \times 10^{21}$ cm$^{-3}$ (device B). Here, we used the Hall factor $r = 0.6$ elucidated for $Sn_{1-x}In_xTe$ in Ref. (22). The $T_c$ data obtained in both devices reasonably follow the trend reported for bulk single crystals[22] as shown in Figure 4; note that the red horizontal bars in Fig. 4 do *not* represent error bars but mark the difference between $p$ and $x$, the agreement of which is reasonably good.

The obtained transport data indicate that the $Sn_{1-x}In_xTe$ nanoplates possess essentially the same superconducting properties as those of large bulk single crystals grown by the same vapor-transported method.[22] Therefore, one may expect that the $Sn_{1-x}In_xTe$ nanoplates grown by the vapor-transported method can be widely used for the studies of superconducting properties of $Sn_{1-x}In_xTe$ and the search for Majorana fermions by employing advanced nanodevice fabrication techniques.

**CONCLUSION**

We developed a simple and clean technique to synthesize superconducting $Sn_{1-x}In_xTe$ nanoplates on Si substrates in quartz glass tube by vapor-transport method with neither transfer agent nor catalyst. We confirmed that $T_c$ and carrier density measured in the nanoplate samples are consistent with those of bulk single crystals grown by the same vapor-transport method, besides the fact that a higher In content and a higher $T_c$ is attainable in the nanoplates. Furthermore, the thickness can be tuned with the growth time, and one can obtain nanoplates with the thickness of around 50 nm, which is convenient for nanodevice fabrication. The superconducting $Sn_{1-x}In_xTe$ nanoplates made available with the present growth technique will foster the experimental research of topological superconductivity and Majorana fermions using nanodevices and contribute to future quantum technologies.

**ASSOCIATED CONTENT**

**Supporting Information**

EMPA qualitative analysis spectra of superconducting $Sn_{1-x}In_xTe$ nanoplate and EPMA elemental maps to show homogeneity of elements in another vapor-grown $Sn_{1-x}In_xTe$ nanoplate. These materials are available free of charge *via* the Internet at http://pubs.acs.org.

## AUTHOR INFORMATION

**Corresponding Author**

*E-mail: y_ando@sanken.osaka-u.ac.jp.

**Author Contributions**

S.S. conceived the growth condition of nanocrystals, fabricated nanodevices, performed all the measurements, and wrote the manuscript with inputs from Y.A. who conceived and supervised the project.

**Notes**

The authors declare no competing financial interest.

## ACKNOWLEDGMENTS

We thank R. Sato for assistance in crystal growths, M. Novak for useful advice on crystal growth, F. Yang, Y. Ohno, and K. Matsumoto for useful advice on nanofabrication, and A. A. Taskin for helpful discussions. We acknowledge Nanotechnology Open Facilities (NOF) and Center of Innovation (COI) programs for nanofabrication facilities and Comprehensive Analysis Center at ISIR, Osaka University, for the EPMA machine. This work was supported by JSPS (KAKENHI 25220708), MEXT (Innovative Area "Topological Quantum Phenomena"

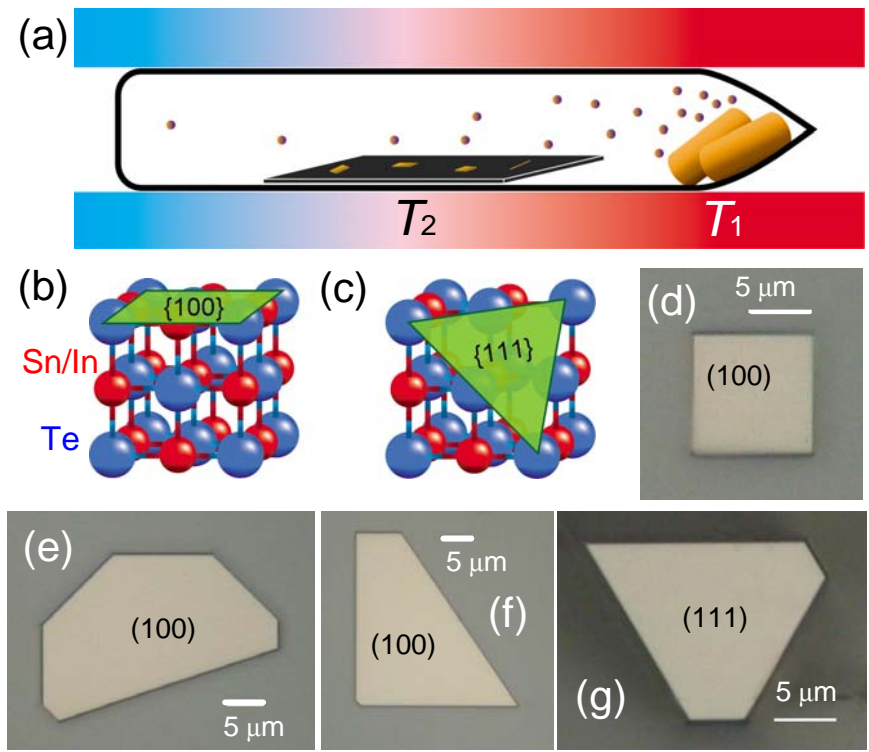

**Figure 1.** Synthesis of $Sn_{1-x}In_xTe$ nanoplates. (a) Schematic of vapor-transport growth method in sealed evacuated quartz glass tube. The source material is placed at $T_1$ (~600 °C) in a 1.4 °C/cm temperature gradient; the temperature of the Si substrate located next to the source material is $T_2$ ($T_1 > T_2$). Transport of SnTe molecules and In atoms inside the quartz glass tube is driven by a temperature gradient and the corresponding gradient in the SVP. (b) {100} facet of $Sn_{1-x}In_xTe$. (c) {111} facet of $Sn_{1-x}In_xTe$. (d, e, f) Optical microscope images of (100) nanoplates. (g) Optical microscope image of a (111) nanoplate. Scale bars are 5 µm.

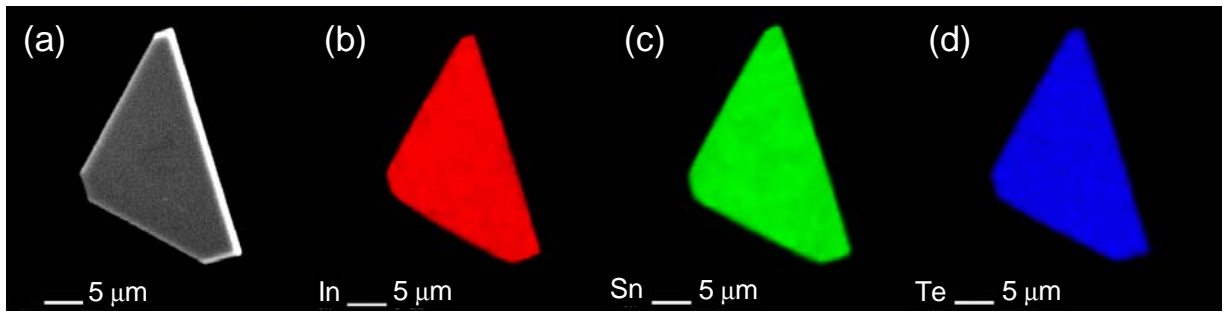

**Figure 2.** Homogeneity of the constituent elements in the $Sn_{1-x}In_xTe$ nanoplate. (a) Secondary electron image of a (100) nanoplate. EPMA elemental maps for (b) In, (c) Sn, and (d) Te. Scale bars are 5 µm.

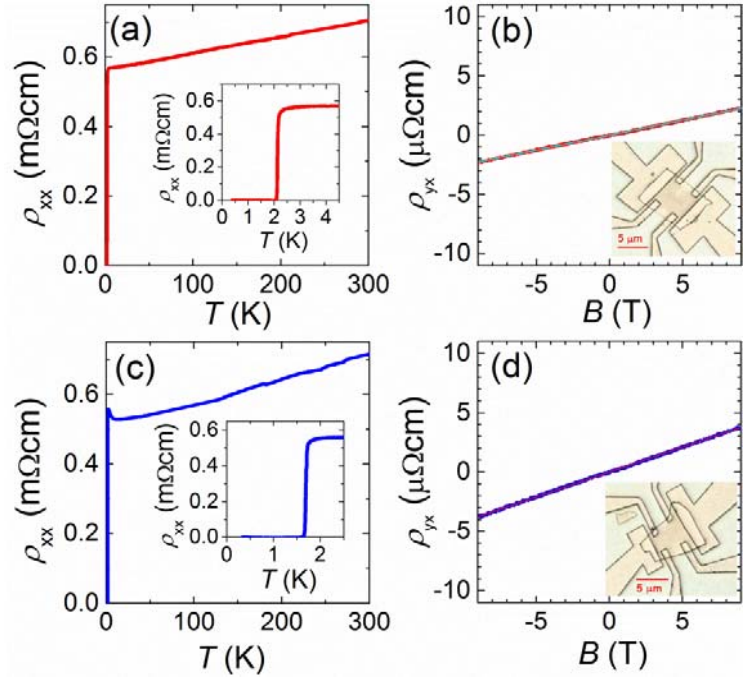

**Figure 3.** Transport properties of the $Sn_{1-x}In_xTe$ nanoplates. (a) Temperature dependence of the resistivity $\rho_{xx}$ of the device A ($x = 0.110$, 57-nm thick). Inset shows a magnified view near the superconducting transition at 2.20 K (mid-point). (b) Magnetic-field dependence of the Hall resistivity $\rho_{yx}(B)$ for the device A. (c) Temperature dependence of the resistivity $\rho_{xx}$ of the device B ($x = 0.061$, 87-nm thick). Inset shows a magnified view near the superconducting transition at 1.75 K. (d) Magnetic-field dependence of the Hall resistivity $\rho_{yx}(B)$ for the device B. In (b) and (d), the solid line shows a linear fitting to the data.

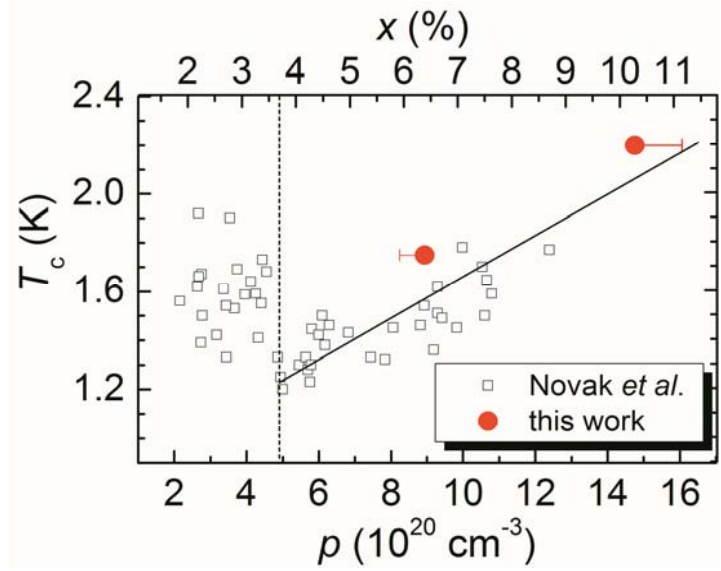

**Figure 4.** $T_c$ vs $p$ plot with the upper horizontal axis showing In content $x$, using the relation between $p$ and $x$ determined in Ref. (22). Open squares represent the data taken from Ref. (22). Closed circles show the data of $T_c$ vs $p$ for $Sn_{1-x}In_xTe$ nanoplates measured in the present work (devices A and B). The red horizontal bar represents the difference between the hole density $p$ obtained from the slope of $\rho_{yx}$ and the In content estimated from the EPMA analysis.

**Supporting Information for**

# Superconducting $Sn_{1-x}In_xTe$ Nanoplates

*Satoshi Sasaki and Yoichi Ando\**

Institute of Scientific and Industrial Research, Osaka University, Ibaraki, Osaka 567-0047, Japan

Corresponding Author

*Y. Ando, E-mail: y_ando@sanken.osaka-u.ac.jp.

1. EPMA qualitative analysis spectra of a superconducting $Sn_{1-x}In_xTe$ nanoplate

To test the high quality of $Sn_{1-x}In_xTe$ nanoplates, we performed electron-probe microanalyzer (EPMA) qualitative analysis. Figure S1 shows an example of the EPMA spectra obtained from such an analysis performed on the superconducting $Sn_{1-x}In_xTe$ nanoplate fabricated to become device A (shown in Figure 3b inset). We confirmed the presence of the constituent elements (Sn, In, and Te) and the absence of other elements. Furthermore, from the height of the In peak, one can quantitatively analyze the In content in the sample.

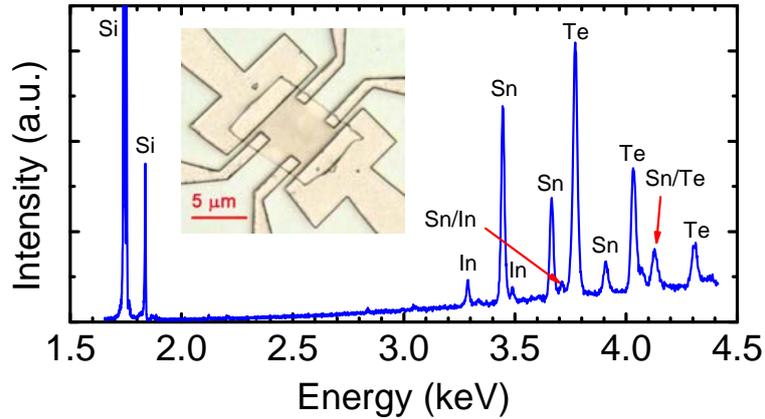

**Figure S1.** EPMA spectra. The characteristic X-ray spectra of the device A (displayed in the inset) document the presence of only Sn, In, Te (constituent elements of $Sn_{1-x}In_xTe$ nanoplate), and Si (element of the substrate). Some of the peaks can be assigned to two elements because of the overlaps in their energy spectra (indicated by the arrows).

2. Homogeneity of constituent elements of another $Sn_{1-x}In_xTe$ nanoplate.

We performed the EPMA elemental mapping of $Sn_{1-x}In_xTe$ nanoplates. The result on one of the samples is shown in the main text. Here the result on another sample is shown to demonstrate the reproducibility of high homogeneity of constituent elements inside the nanoplates.

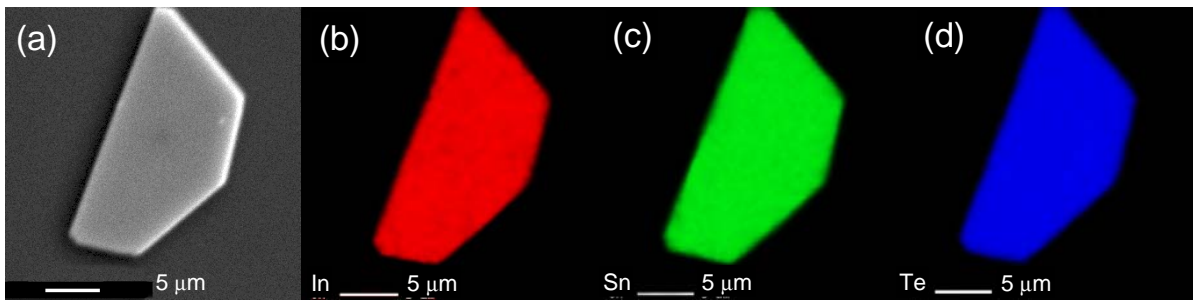

**Figure S2.** Homogeneity of constituent elements in the $Sn_{1-x}In_xTe$ nanoplate. (a) Secondary electron image of a (100) nanoplate. EPMA elemental maps for (b) In, (c) Sn, and (d) Te. Scale bars are 5 μm.

2